\documentclass{ws-ijmpa}

\begin{document}

\markboth{G. Brooijmans}
{Searches for New Physics}

%
\catchline{}{}{}{}{}
%

\title{SEARCHES FOR NEW PHYSICS}

\author{\footnotesize GUSTAAF BROOIJMANS\footnote{email:gusbroo@fnal.gov}}

\address{Physics Department, Columbia University\\
New York, NY 10027, USA}

\maketitle


\begin{abstract}
Current experimental limits for new physics beyond the Standard Model and hints
for deviations from Standard Model expectations will be reviewed, highlighting
recent results.  Possible signals that will be discussed include Higgs bosons,
supersymmetric particles, large extra dimensions, new gauge bosons, dynamical
symmetry breaking, muon g - 2, rare decays and  lepton flavor violation.  The
discovery potential of the LHC and ILC will be presented, and the impact of
discovery on answering fundamental questions of physics will be assessed.
 
\end{abstract}

\section{Introduction}

This talk is will cover direct and indirect searches for new physics, 
as well as prospects for the future.  Looking at the parallel session
agendas reveals the large number of topics to cover: there were 31 talks in 
the ``Direct Searches'' session, 10 in ``Muon $g-2$, Lepton Flavor Violation 
and Electric Dipole Moments'', 9 in ``LHC-LC Comparison'' and 43 in non-top
``Heavy Flavor Physics''.  This overview can therefore not be exhaustive, and 
the author apologizes to all those whose work is not shown here.

\section{The Standard Model and Its Caveats}
\label{sec:ques}

It is sometimes useful to formulate the standard model of particle physics (SM)
in words to expose its strengths and weaknesses:
\begin{itemize}
\item Matter is built of spin 1/2 particles that interact by exchanging three
 different kinds of spin 1 particles corresponding to three different (gauge)
 interactions.
\item There are three generations of matter particles.
\item The four different matter particles in each generation have different 
 combinations of (quantified) charges characterizing their couplings to the
 interaction bosons.
\item The matter fermions and the weak interatcion bosons have ``mass''.  This 
 is called electroweak symmetry breaking (EWSB).
\item Gravitation is presumably mediated by spin 2 gravitons.
\item There appear to be three macroscopic space dimensions.
\end{itemize}
The last two items in the list are not strictly speaking part of the 
standard model, but are generally implicitely assumed.

These very much simplified statements raise a large number of questions.  
At this particular time, the author considers the following ones to be 
particularly fundamental:
\begin{itemize}
\item What exactly {\em is} (weak iso)spin?  Or color?  Or electric charge?
 Why are they quantified?
\item Are there only three generations?  If so, why?
\item Why is there no matter that doesn't interact weakly?  Or why, for example,
 are there no neutral, colored fermions?
\item What is mass?  Is it quantified?
\item How does all of this reconcile with gravitation?  How many space-time
 dimensions are there really?
\item Is ``our universe'' the unique solution?
\end{itemize}
Many other questions can be asked (see for example Ref.~\cite{kane1}).  
Answering any of these unambiguously at a fundamental level would be a major 
breakthrough in physics.

\section{Mass}

Among the fundamental questions cited above, the one we think we have a 
handle on is mass.  The addition of a na\"ive $M^2WW$ mass term to 
generate the gauge boson masses (luckily) not only breaks gauge invariance,
but also destroys the renormalizability of the standard model.  In fact, 
at high energy ($\sqrt{s} \approx 1.7\ TeV$), $W_LW_L$ scattering violates
the Froissard bound.  And elegant solution to this problem is provided by
the Higgs mechanism: the ``standard model Higgs''generates both boson and
fermion masses, and ``restores'' unitarity if $m_H \lesssim 1\ TeV/c^2$.  

Since in the standard model the couplings of the Higgs boson to all 
particles is known, its mass can be inferred from precision measurements
that are sensitive to processes in which Higgs bosons contribute at the
one- or multiple loop level.  This can be seen in 
Figure~\ref{fig:hm}~\cite{higgsewwg}:
\begin{figure}
\centerline{\psfig{file=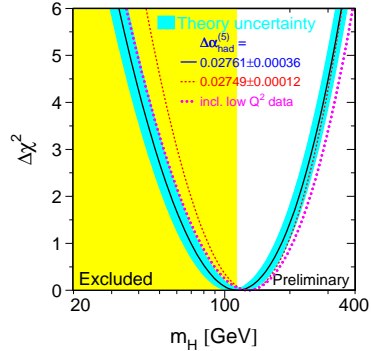,width=5cm}}
\caption{\label{fig:hm} Higgs boson mass inferred from precision measurements.  
The yellow
shaded area is excluded based on direct searches at LEP2.}
\end{figure}
the yellow shaded area is excluded based on direct searches at LEP2 giving
$m_H > 114.4\ GeV/c^2$ at 95 \% C.L., while the curve is the result of the fit
to precision data.  It should be noted that this is very sensitive to the 
measured values of the top quark and W boson mass: the recent increase 
in the top quark mass~\cite{d0top} by $4\ GeV/c^2$ has shifted the best fit value
up by $18\ GeV/c^2$.  The best fit value is now $m_H = 114\ GeV/c^2$.

\begin{figure}
\centerline{\psfig{file=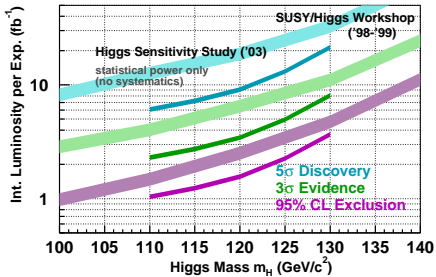,width=6cm}}
\caption{\label{fig:higgstev} Tevatron projected sensitivity for Higgs boson 
discovery as a function of Higgs boson mass and integrated luminosity.  The 
thinner lines are from the 2003 Higgs Sensitivity Study, 
the thicker from the 1998-1999 SUSY/Higgs Workshop.  Current projections estimate
that each experiment will receive between $4$ and $8\ fb^{-1}$ by 2009.}
\end{figure}

\begin{figure}
\centerline{\psfig{file=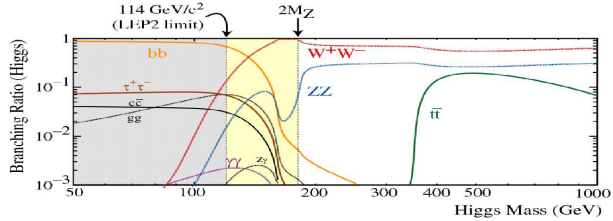,width=8cm}}
\caption{\label{fig:higgsbr} Standard model Higgs boson branching 
fractions as a function of its mass.}
\end{figure}

If a standard model Higgs boson exists (here standard model denotes that it
is the sole source for both the boson and fermion masses), prospects for 
establishing its existence before the end of the decade are excellent.  If 
it is relatively light, some signal should be visible at the Tevatron~\cite{htev} before
the LHC produces physics results, as can be seen in Figure~\ref{fig:higgstev}.
It should be noted that in this study only the channels in which a Higgs
boson is produced in association with a $W$ or $Z$ boson and decays to
$b \overline{b}$ have been considered, while at $m_H = 120\ GeV/c^2$, 
Figure~\ref{fig:higgsbr} shows that the 
branching fraction to a pair of $W$ bosons is already quite substantial.
This promising channel is actively pursued by both the CDF and D\O\ collaborations.

\begin{figure}
\centerline{\psfig{file=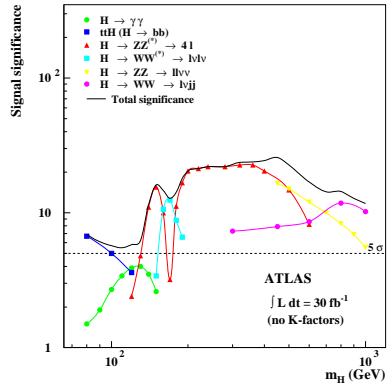,width=5cm}}
\caption{\label{fig:higgslhc} Sensitivity of the ATLAS 
experiment for 
standard model Higgs boson discovery over the full mass range for an 
integrated luminosity of 30 $fb^{-1}$, corresponding to three years 
of low luminosity running.  The differently colored lines correspond to 
different channels.  Note that for low masses, the rare $H \rightarrow 
\gamma \gamma$ channel is critical.}
\end{figure}

In case the standard model Higgs exists, but is too heavy to be seen at 
the Tevatron, discovery at the LHC within a few years of running is just
about certain over the full mass range $100\ GeV/c^2 \leq m_H \leq 1\ TeV/c^2$, 
as shown in Figure~\ref{fig:higgslhc}~\cite{atlastdr}.

Discovery of a Higgs boson will be followed by accurate measurement 
of its properties.  Determining that it is indeed the source of all 
particle masses (i.e. its coupling to the particles), 
its spin, its self-coupling, etc. will then be key in 
increasing our understanding of mass as a fundamental property of 
matter.  Measurement of the couplings to the heavier fermions and the 
weak bosons 
can be done through determination of production and decay rates, the
former mainly in associated production processes, and the latter limited 
to decay channels with reasonable rate for a given mass.  At the LHC, a
few years of low luminosity running can yield measurements at the 10-50\%
level~\cite{duehrssen}, with best precision on the coupling to $W$ bosons.

The International Linear Collider will then be an ideal place to study
the Higgs boson further, and coupling strengths could be measured with
precision of better than 1\% depending on the Higgs boson mass~\cite{ilc1}.
The Higgs spin can be verified by measuring the production cross-section 
as a function of center-of-mass energy~\cite{dova}, although it will 
probably already have been determined from LHC data (the detection of 
$H \rightarrow \gamma \gamma$ excludes spin 1 as a possibility for example).

\section{Models of New Physics}

While the discovery of a standard model Higgs boson would represent a 
significant step forward, it really doesn't tell us what mass
is.  The question can then be stated as three new, separate questions:
why are the Yukawa couplings what they are;  why is $\mu^{2}$ in the 
Higgs potential negative; and what is the link to gravity? 

The Higgs mechanism also introduces a new set of problems (or benefits,
depending on one's point of view).
The Higgs boson mass is ``naturally'' the energy scale at which some new
physics manifests itself, so if we have a standard model Higgs, that's 
about $200\ GeV$.  There are two theoretical approaches to accomodate this:
fixing by addition (introduction of new particles and - sometimes - 
interactions with masses 
${\cal O}(200-1000\ GeV/c^2)$ to stabilize the Higgs mass), 
or fixing by subtraction (no Higgs boson).

\subsection{Low Scale Supersymmetry}

Supersymmetry (SUSY) is certainly the most popular of the existing models, for good 
reason.  In supersymmetry, for each boson there is an associated 
fermion, and vice-versa.  All quantum numbers, except spin, and all 
couplings are the same for the so-called ``superpartners'', but their 
masses appear to be different since none of these ``sparticles'' have been
observed.  This is SUSY breaking.  As will be shown, low-scale SUSY has
a number of attractive features, but one major disadvantage, at least 
in the author's mind, is that it trivializes spin.  So what are 
the big advantages of SUSY?  First, the fermionic and bosonic 
loop corrections to the Higgs mass cancel each other, so the 
Higgs boson mass is naturally of the same order as the SUSY mass
scale.  Second, with the added particles, gauge coupling unification
is much improved w.r.t. the standard model~\cite{unif1}, in the sense that
all three couplings converge in one point just above $10^{16}\ GeV$.
Third, SUSY explains EWSB, as will be shown below.

\begin{figure}
\centerline{\psfig{file=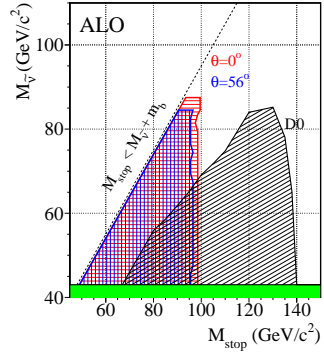,width=5cm}}
\caption{\label{fig:ld0stop} LEP and D\O\ limits on stop and sneutrino
masses assuming the stop decays to $b l \tilde{\nu}$ 100 \% of the
time.}
\end{figure}

\begin{figure}
\centerline{\psfig{file=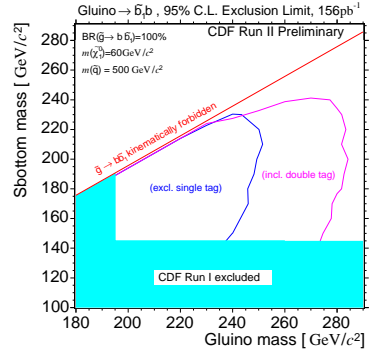,width=5cm}}
\caption{\label{fig:cdfsbot} CDF limits on gluino and sbottom masses
assuming gluinos decay to a sbottom and bottom 100\% of the time.}
\end{figure}

The minimal supersymmetric model means a minimal number of additional 
particles, but also a minimal number of constraints.  This introduces
105 new parameters (sparticle masses, mixing angles, \ldots).  Some 
searches for superpartners are done in this context, with a very small
number of additional assumptions.  Typically these are that the Lightest
Supersymmetric Particle\footnote{Along with SUSY, a new symmetry called 
$R$-parity has been introduced.  Under this symmetry, standard model 
particles have eigenvalue $1$ while sparticles have eigenvalue $-1$, so 
that the LSP is stable if $R$-parity is conserved.},
or LSP, is the lightest neutralino (partners
of the neutral bosons mix), and an assumption on the branching fraction 
for the process studied.  Figures~\ref{fig:ld0stop} and \ref{fig:cdfsbot}
show examples of these performed by the LEP collaborations, CDF and 
D\O\ .

\begin{figure}
\centerline{\psfig{file=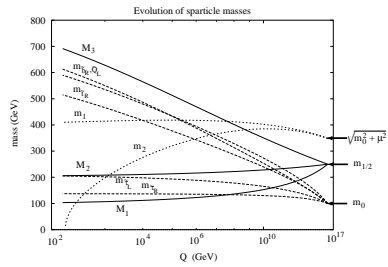,width=5cm}}
\caption{\label{fig:sugrarun} Running of sparticle masses in minimal 
supergravity (mSUGRA).  Note $m_2$ ($\mu^2$), which is driven negative at 
low energy.} 
\end{figure}

Various SUSY breaking models exist,
with very different phenomenological signatures, but one general 
feature is that they demonstrate that supersymmetry can explain 
electroweak symmetry breaking.  This is
illustrated in Figure~\ref{fig:sugrarun}~\cite{susy1}: in supersymmetry, when 
the renormalization group equations are used to run the couplings 
down to the electroweak scale, the $\mu^2$ term in the Higgs potential
is naturally driven negative, triggering EWSB.

\begin{figure}
\centerline{\psfig{file=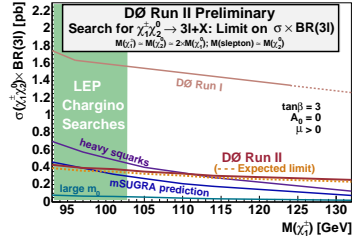,width=5cm}}
\caption{\label{fig:d0tril} D\O\ Run II limit on associated chargino-neutralino
production and their decay to trileptons based on 147-249 $pb^{-1}$ of data.
The red curve denoted ``D\O\ Run II'' shows the cross-section limit from the 
analysis, while the purple, blue and green curves represent cross-section
predictions in various models.  Chargino masses with a predicted cross-section 
higher than the cross-section limit are excluded.}
\end{figure}

In one of the models, supergravity (SUGRA), SUSY breaking is transmitted 
from a hidden sector through
gravity.  This reduces the number of free parameters to five in the minimal
version of the model (mSUGRA).  $R$-party is generally assumed to be conserved. This is 
the SUSY framework in which most searches are conducted.  At the Tevatron, the 
golden signature for mSUGRA is the trilepton signature from associated 
production of $\chi^{\pm}_1 \chi^0_2$ (the lightest chargino and next-to-lightest
neutralino, respectively) and their decay through virtual sleptons and gauge 
bosons.  The recent result from D\O\ shown in Figure~\ref{fig:d0tril} shows
that the experiment's sensitivity is close to exceeding the range excluded by
LEP.

\begin{figure}
\centerline{\psfig{file=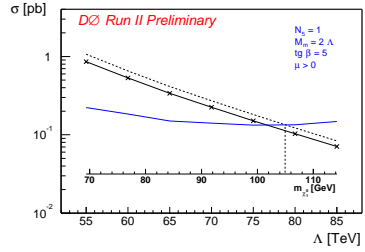,width=5cm}}
\caption{\label{fig:d0ggmet} D\O\ Run II limit on GMSB with a bino LSP.
The parameter $\Lambda$ is the supersymmetry scale in these types of 
models; the corresponding neutralino mass is also given.}
\end{figure}

Another extensively studied model of SUSY breaking is gauge mediated SUSY
breaking, or GMSB.  In this scenario, SUSY is also broken in a hidden sector,
but the SUSY breaking messengers participate in standard model gauge interactions,
and superpartner masses are therefore proportional to gauge boson couplings. 
The LSP is a very light (sub-$eV$) gravitino and the phenomenology is 
driven by the nature of the next-to-lightest supersymmetric particle (NLSP), 
which decays to its partner and the gravitino.  This 
model's major support comes from the famous 
$e e \gamma \gamma \slash$\hspace{-0.25cm}$E_T$ candidate
event detected by CDF in Run I of the Tevatron~\cite{cdfeegg}.  In most 
instantiations of this model, the NLSP is either a slepton or the bino, 
partner of the unmixed $B$ field in electroweak theory.  At hadron 
colliders, the latter thus 
leads to signatures with two high-energy photons and missing transverse
energy. The current best limit in this scenario is the recent result from 
D\O\ shown in Figure~\ref{fig:d0ggmet}~\cite{d0ggmet}.

\begin{figure}
\centerline{\psfig{file=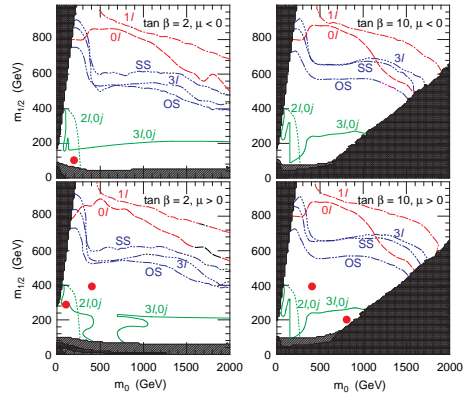,width=6cm}}
\caption{\label{fig:atlsusy} Supersymmetry reach of the Atlas experiment in 
one year of low luminosity running (10 $fb^{-1}$) for various signatures, 
including lepton plus jets (denoted $1l$), 
jets plus missing transverse energy ($0l$), trileptons
($3l$), same- and opposite sign dileptons (SS and OS), etc.  The four different
plots correspond to different choices for two of the other mSUGRA parameters.}
\end{figure}

In a supersymmetric world there are two Higgs doublets, and five physical Higgs 
bosons (versus one and one in the standard model).  There are two charged Higgses
($H^\pm$), two $CP$-even Higgses ($h$ and $H$) and one $CP$-odd Higgs ($A$).
At the LHC, at least one of these can always be seen, although if $m_A$ is large
for much of the parameter space this Higgs is not distinguishable from the 
standard model Higgs.  Luckily, in this region, other supersymmetric particles
are often visible.  Indeed, if the supersymmetry scale is less than about 
$1\ TeV$, supersymmetry should manifest itself through a plethora of 
signatures at the LHC.  This is illustrated in 
Figure~\ref{fig:atlsusy}~\cite{atlastdr}
where ATLAS' sensitivity to mSUGRA is illustrated in various channels
for one year of running at low luminosity (10 $fb^{-1}$).  

\begin{figure}
\centerline{\psfig{file=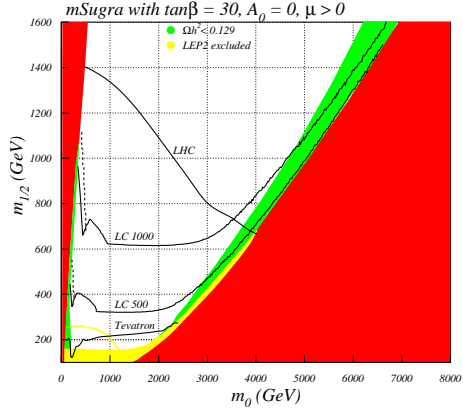,width=6cm}}
\caption{\label{fig:susylhcilc} Supersymmetry reach of the Tevatron, LHC 
and ILC with two different center-of-mass energies as a function of the 
mass parameters of mSugra for a particular choice of the other parameters.
The red regions are theoretically excluded, while the yellow area has been 
excluded by LEP2.  The green bands are the regions preferred by the WMAP 
dark matter measurement: at large $m_{1/2}$ but low $m_0$ the so-called 
coannihilation region, and at large $m_0$ the focus points where the 
LSP has a large higgsino component.}
\end{figure}

Since in supersymmetry the LSP is usually expected to be stable, experiments will 
detect cascade decays of heavier particles, with the LSP detected through
the presence of missing (transverse) energy.  
Therefore invariant masses cannot be reconstructed directly, and to determine 
sparticle masses the endpoints of kinematic distributions are used.
If enough superpartners are accessible,
then the pattern of sparticle masses can be measured and used to 
try to deduce the mechanism of SUSY breaking.  At the LHC, it should be 
possible to measure the masses of the accessible fermion partners with 
$5-10\ GeV/c^2$  accuracy, and this could be improved to $1\ GeV/c^2$ or better
at the ILC~\cite{lhclc}.  Of course, this is very dependent on the actual
spartner masses and the situation could be less favorable.  Studies to 
determine the reach of both the LHC and ILC have been made, with one example
illustrated in Figure~\ref{fig:susylhcilc}~\cite{baer1}.
It is interesting to note that a $1\ TeV$ ILC has sensitivity in the 
focus point area at large $m_0$ which is out of reach of the LHC.

In addition to direct searches for on-shell production of supersymmetric 
particles, rare decays and precision measurements can yield interesting
data on SUSY parameters.  Rare decays are processes in which the tree-level
diagram is forbidden, for example because it is a flavor changing neutral 
current (FCNC).  At the one loop level these often still involve a weak
process, contributing further to keeping the branching fraction low.
If backgrounds from other processes are small, these can provide a means 
of probing physics at one or two loops.  Other processes are measured with
stunning precision, and can be calculated to similar accuracy, such that any 
(lack of) deviation can be interpreted in terms of new physics parameters.

\begin{figure}
\centerline{\psfig{file=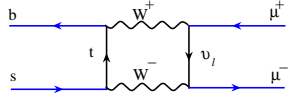,width=5cm}}
\caption{\label{fig:bsm1}  One of the standard model box diagrams 
contributing to $B_s \rightarrow \mu^+ \mu^-$ decays.}
\end{figure}

\begin{figure}
\centerline{\psfig{file=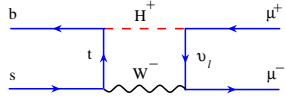,width=5cm}}
\caption{\label{fig:bsm2}  A box diagram
contributing to $B_s \rightarrow \mu^+ \mu^-$ decays in 
two Higgs doublet models like for example SUSY.}
\end{figure}

A good example is the search for the decay $B_s \rightarrow \mu^+ \mu^-$,
for which the standard model branching fraction is expected to 
be~\cite{buch} $3.8 \times 10^{-9}$.  An example of a standard model 
diagram is given in Figure~\ref{fig:bsm1}.
SUSY diagrams like the one illustrated in Figure~\ref{fig:bsm2}
could increase this value by up to three orders of magnitude.  
The Tevatron is currently the only copious source of $B_s$ mesons and recent 
results from CDF~\cite{cdfbsm} ($BR(B_s \rightarrow \mu^+ \mu^-) < 5.8 \times 10^{-7} $) and 
D\O\ \cite{d0bsm} ($BR(B_s \rightarrow \mu^+ \mu^-) < 5.0 \times 10^{-7} $) have been 
combined by M. Herndon~\cite{herndon} to yield\footnote{Since this combination was
made, the D\O\ result shifted from $BR(B_s \rightarrow \mu^+ \mu^-) < 4.6 \times 10^{-7}$
to $BR(B_s \rightarrow \mu^+ \mu^-) < 5.0 \times 10^{-7}$ at 95\% C.L. and the 
combined value is therefore expected to be slightly higher than quoted in the text.} 
$BR(B_s \rightarrow \mu^+ \mu^-) < 2.7 \times 10^{-7}$ at 95\% C.L., an
improvement of an order of magnitude over the earlier result.  

Similarly, BABAR searches for $B_d \rightarrow l^+ l^-$ decays.  In $120\ fb^{-1}$ of
on- and off-resonance data~\cite{babar} they set the following limits:
$BR(B_d \rightarrow e^+ e^-) < 6.1 \times 10^{-8}$ (standard model 
expectation is $1.9 \times 10^{-15}$), 
$BR(B_d \rightarrow \mu^+ \mu^-) < 8.3 \times 10^{-8}$ ($8.0 \times 10^{-11}$),
$BR(B_d \rightarrow e^{\pm} \mu^{\mp}) < 18 \times 10^{-8}$ ($0$).  This 
allows them to put a limit on the Higgs mass in the minimally constrained 
supersymmetric standard model at $m_H > 138\ GeV$ at 90\% C.L. for 
$\tan{\beta} = 60$.  Analogous analyses are done with the measurement of
$B \rightarrow X_s \gamma$ or rare tau decays.

At Brookhaven, experiment E949 has recently observed another 
$K^+ \rightarrow \pi^+ \nu \overline{\nu}$ candidate event, bringing their
branching ratio measurement~\cite{e949} to 
$BR(K^+ \rightarrow \pi^+ \nu \overline{\nu}) = 1.47 ^{+1.30}_{-0.89} \times 10^{-10}$,
whereas the current standard model estimate~\cite{deandrea} is 
$BR^{SM} = 8.18 \pm 1.22 \times 10^{-11}$.  The same authors~\cite{deandrea} calculate that the
SUSY contribution without R-parity violation can account for a maximum of 50\% of the
standard model value when taking into account the current SUSY bounds, and use this
to set limits on some R-parity violating couplings that are the 
most stringent to date.

The measurement of the anomalous magnetic moment of the muon is a difficult, but 
well-understood experiment, now done to a stunning precision of 0.5 parts per 
million~\cite{g-2}.  
This means it is sensitive to two-loop corrections, and thus has great potential to 
see effects from heavy new particles.  The theoretical value is correspondingly well
known, to about 0.7 ppm.  There is some variation in the calculations, however, and
there are always two results, depending on which data is used as input for the 
corrections due to hadronic vacuum polarization.  
\begin{figure}
\centerline{\psfig{file=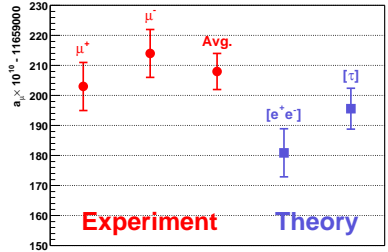,width=5cm}}
\caption{\label{fig:g-2}  Current experimental and theoretical values for
the muon anomalous magnetic moment, from the E821 experiment }
\end{figure}
Figure~\ref{fig:g-2} shows the experimental values for both muon charges and their 
average, as well as a theoretical calculation (with two results as discussed above).
The discrepancy between experiment and theory is two to three sigma, depending
on which theoretical calculation is chosen, but the difference is always in the 
same direction.  It should be noted that this discrepancy is larger than the effect
of weak interactions by 30\% , and it is therefore difficult to imagine that it's
due to new physics in its entirety~\cite{detroc}.  This situation makes it possible
to set very stringent constraints on any new physics effects that would lead the 
theoretical result to pull even further from the experimental measurement.  An 
example is given in Reference~\cite{heine1}, where it is shown that a negative 
value of the parameter $\mu$ in the minimal supersymmetric standard model is now strongly 
disfavored.

\subsection{Technicolor}

While low-scale supersymmetry presents the advantages of solving the hierarchy problem,
improving gauge coupling unification, and explaining mass, its prediction of the existence
of elementary scalars and the fact that no evidence for SUSY has as yet been found, 
continue to inspire the development of alternate models.  One of these is technicolor, a QCD-inspired, 
strongly coupled theory.  In technicolor, the hierarchy between the electroweak and
unification scales is explained as a confinement phenomenon of a new interaction, in 
analogy to the pion mass in QCD.  Technifermions are bound into technihadrons, but 
there are no fundamental scalars.  The strong technicolor coupling makes it difficult 
to satisfy the constraints from precision data, forcing increased complexity into the
model, and we now have topcolor-assisted walking technicolor.
Another unfortunate side effect of the strong coupling is that it makes calculations 
and therefore predictions
difficult.  At hadron colliders, it is thought that vector technimesons are most 
likely to be produced, and these then decay to technipions and longitudinally
polarized vector bosons.  The technipions act a like the Higgs boson and 
couple to mass, and are therefore expected to decay primarily to heavy quarks.
Searches for technicolor at colliders are performed in the same final states as
Higgs searches ($Wb\overline{b}$ at the Tevatron for example) and in dilepton
final states~\cite{lorenzo}.

\subsection{Extra Dimensions}

In the late 1990s, it was proposed~\cite{add} that the extra dimensions predicted 
by string theory could be quite large, and accessible at colliders.  In the original ADD
model, standard model particles are confined to a 3-brane with gravity propagating in 
more dimensions.  The hierarchy problem\footnote{The large difference between the 
electroweak and Planck scales is commonly referred to as the hierarchy problem.} 
is solved by bringing down the fundamental Planck scale, 
which only appears high in three dimensions.  In this model, there are two main types 
of signatures: on the one hand, interference from Kaluza-Klein graviton 
excitations~\footnote{Particles that propagate in compactified extra dimensions have
quantified momenta in those dimensions, which appear as mass to the observer.  Many of 
these particles, which have different momenta in the extra dimensions, form a
so-called ``tower'' of Kaluza-Klein excitations.}
can visibly affect the high energy and angular behavior of standard model processes.  
This is because the graviton couples to the energy-momentum tensor, and at higher energies 
more and more excitations are accessible while the standard model cross-section typically 
falls fast.  A second class of signatures comes from production of on-shell Kaluza-Klein 
graviton excitations, which then disappear back into the extra dimensions, 
leaving a missing energy signature.
The current best limit on large extra dimensions in the ADD model was presented at 
this conference~\cite{ryan} and comes from the D\O\ experiment: 
$M_S > 1.43\ TeV$ at 95\% C.L. in the GRW convention.  The LHC is expected 
to be sensitive up to about $9\ TeV$~\cite{lhced}, depending on the number of dimensions.

\subsection{Resonances}

Many new physics models predict the existence of high mass resonances.  This includes
the warped extra dimensions model of Randall and Sundrum~\cite{rs} and its variations,
and Little Higgs and other models which exhibit extended group structures.  The 
resonances themselves can be graviton excitations, gauge boson Kaluza-Klein 
excitations, or extra $W', Z'$ gauge bosons with various coupling strengths and widths.
Experimentally, analyses are done by final state and a single analysis is used to
constrain many models.
\begin{figure}
\centerline{\psfig{file=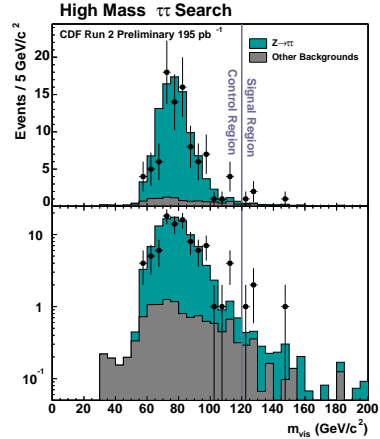,width=5cm}}
\caption{\label{fig:cdfdtau} Search for extra gauge bosons in the ditau channel at
the CDF experiment: the plots show the visible ditau invariant mass for all 
candidate events, in both the control region below $m_{\tau \tau} < 120\ GeV/c^2$
and in the signal region above.  This allows CDF to put a limit on a sequential 
$Z'$ boson in this decay channel at approximatively $m_{Z'} \leq 400\ GeV/c^2$ at
95\% C.L.}
\end{figure}
\begin{figure}
\centerline{\psfig{file=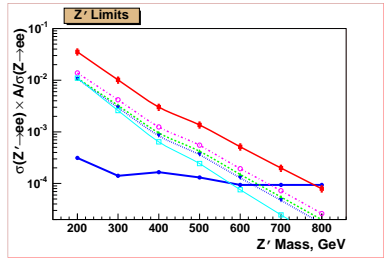,width=5cm}}
\caption{\label{fig:d0zpr} Limits on extra gauge bosons from the D\O\ experiment.
The blue line is the experimentally excluded cross-section, the red line the 
predicted cross section for a sequential $Z'$ boson with couplings identical to the 
standard model $Z$ boson, and the other lines represent production cross-sections 
for $Z'$ bosons in various supersymmetric $E_6$ models.}
\end{figure}
Figure~\ref{fig:cdfdtau} shows the invariant mass spectrum for ditau events in the
CDF experiment.  This distribution allows CDF to put a limit on a sequential 
$Z'$ boson in this decay channel at approximativeky $m_{Z'} \leq 400\ GeV/c^2$ at
95\% C.L.  In Figure~\ref{fig:d0zpr}, the cross-section limit obtained by the D\O\
experiment in its search for $Z'$ bosons decaying to $e^+ e^-$ is shown together
with production cross-sections from various models.  The limit derived for a 
sequential $Z'$ is $m_{Z'} \leq 800\ GeV/c^2$ at 95\% C.L.

If a new high-mass resonance is discovered, the distinction between a spin 1 and 
spin 2 particle can in principle be made by measuring the angle between one of 
the decay leptons and the beam direction in the dilepton center-of-mass 
frame~\cite{all1}, usually called $\theta^*$.  However, if parameters conspire,
this may not be possible at a hadron collider~\cite{rizzo1}, and in any case a
linear collider would probably be necessary to distinguish between similar 
models by exploiting the lineshape~\cite{rizzo2}.

At the LHC, resonances of mass up to almost $6\ TeV/c^2$ could be discovered 
provided the underlying model parameters are favorable.  A linear collider would
obviously be an ideal machine to study such a resonance, provided it is within 
its energy reach.

\subsection{Split Supersymmetry}

N. Arkani-Hamed and S. Dimopoulos recently proposed a model~\cite{ssusy} in which 
they argue that the cosmological fine-tuning problem suggests that fine-tuning 
itself might be an intrinsic part of nature.  This model therefore doesn't address
the electroweak hierarchy problem, but mainly attempts to improve gauge unification 
over the standard model.  This is achieved by keeping the scalars ultraheavy, 
but making the fermionic gauginos light through chiral symmetry.  This 
has been baptised ``split supersymmetry''.  The prediction for the Higgs mass is 
somewhat relaxed compared to low scale supersymmetry (to $m_H \lesssim 150\ GeV/c^2$), 
gauge coupling unification is achieved, the golden trilepton signature is still 
predicted since gauginos are light.
While the model is new, the new feature in its phenomenology,
a long-lived gluino, gives rise to signatures which are already searched for:
an escaping neutral gluino hadron leads to the same monojet plus missing transverse
energy predicted by ADD large extra dimensions, while charged gluino-hadrons would
lead to charged massive particles which are searched for in the context of 
gauge-mediated supersymmetry breaking with a charged, long-lived NLSP.  Results for
both types of searches have been presented at this conference~\cite{ssusyex}.

\section{Direct Probes}

Some of the fundamental question asked in section~\ref{sec:ques} can be investigated
more-or-less directly, independently of a predictive model.  A few examples are given here.

\subsection{Lepton Flavor Violation}

The generational structure of the standard model fermions clearly suggests that 
there is a link between the different generations.  Therefore, lepton flavor 
violation is expected at some scale, and it is in fact seen in the neutrino 
sector over long distances.  Experimentally, lepton flavor violating muon decays
or conversions yield a very sensitive probe to high scale physics.  In the search 
for $\mu \rightarrow e \gamma$ decays, the MEGA experiment has set a branching ratio
limit at $< 1.2 \times 10^{-11}$ at 90\% C.L.~\cite{mega}, with the MEG 
experiment~\cite{meg} 
set to explore values down to $10^{-14}$.  Conversion of muons to electrons in  
in the Coulomb field of a nucleus $\mu N \rightarrow e N$ is a similar process 
which could be more sensitive depending on the physics process of lepton flavor
violation.  For this process, the current best limit is set by the SINDRUM II 
experiment~\cite{sindrum} at $B_{\mu e} < 6.1 \times 10^{-13}$ at 90\% C.L., 
and improvement by three orders of magnitude is expected from the MECO 
experiment~\cite{meco}. 

\subsection{Proton Decay}

If there is indeed unification of quarks and leptons at some energy, then the 
proton is unstable.  In fact, the non-observation of proton decay at this stage is
already putting stringent constraints on a number of unification models, like
minimal SUSY SU(5) GUT.  The current limit on the proton lifetime is 
$\tau > 10^{31} - 10^{33}$ years~\cite{superk} depending on the decay mode 
studied.  In the next generation detector, which will most likely be a megaton water cherenkov
detector, the sensitivity is expected to reach $10^{35}$ years, very close to 
the expectation of $10^{36}$ years from gauge coupling unification~\cite{bour}.

\subsection{Magnetic Monopoles}

\begin{figure}
\centerline{\psfig{file=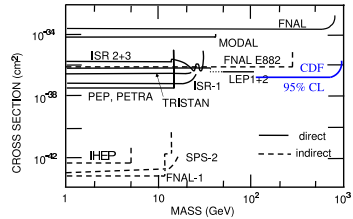,width=5cm}}
\caption{\label{fig:cdfmm} 95\% C.L. limit on magnetic monopole cross-section as 
a function of the particle mass.}
\end{figure}

In 1931 Dirac~\cite{dirac} demonstrated that the existence of even a single 
magnetic monopole would explain electric charge quantization.  In most Grand
Unification Theories (GUTs), magnetic monopoles appear, but with masses of the
order of the unification scale.  It is quite possible that much lighter monopoles
exist and are produced in high energy colliders.  Since magnetic charge is conserved,
they are stable, and can be detected either as magnetic charges trapped in detector 
elements, as done by H1~\cite{h1mm}, or as they travel through the detector.  The latter
method has been exploited by CDF to set the best limit to date on these particles,
as shown in Figure~\ref{fig:cdfmm}.

\section{Model Independent Searches}

In addition to model-driven searches, there is an increase in so-called model-independent
or general searches.  In channels where backgrounds are sufficient small and/or understood
it is indeed possible to achieve discrimination without resorting to sophisticated
analysis of kinematic and topological distributions.  A recent example comes from the 
H1 experiment~\cite{h1gen}, in 
which events are classified exclusively according to their final state, and the scalar
sum of transverse momenta or invariant mass of final state particles 
are compared to expectation for each channel.  These types of analyses have the added
benefit of showing clearly that moderate excesses in a small number of channels are 
to be expected, just as in some channels the observed number of events is smaller
than the prediction.

\section{Current Experimental Hints}

There is currently no conclusive evidence of any physics beyond the standard model,
but there are of course some deviations from expectation.  In addition to those 
described above, there is an interesting excess of events with isolated leptons 
at HERA.  The total number of events with an isolated lepton and large missing transverse
energy is in reasonable agreement with standard model predictions, but when in addition 
to that the recoil system is required to have large transverse momentum, a clear excess is 
seen in H1~\cite{h1il}.  For $p_T^X > 25\ GeV/c$ (where $p_T^X$ is the transverse momentum 
of the recoil hadronic system), 5 and 6 events are observed in the electron 
and muon channels respectively, while $1.8 \pm 0.3$ and $1.7 \pm 0.3$ are expected, a 
2.8 sigma effect.
No excess is seen in those channels in ZEUS, but they observe~\cite{zil} two events in the tau
channel with $0.2 \pm 0.05$ expected.  In a recent article~\cite{dom}, the compatibility 
of the two experimental results is investigated quantitavely with minimal model dependence.
The authors conclude that the most plausible new physics explanation, anomalous tau 
production, has a probability of agreement with observation of a few percent.  
Anomalous single top or $W$ boson production yield probabilities well below one percent.

\section{Answering Fundamental Questions}

Throughout this talk, some of the fundamental questions posed early on have been 
addressed:  
\begin{itemize}
\item Understanding electroweak symmetry breaking, possibly through the discovery 
of supersymmetry, would explain mass.  
\item Understanding (the breaking of) grand unification will tell us about electric
charge, color, and possibly spin.  For this, both direct and indirect searches are
critical to acquire knowledge of both low and high scale processes.
\item Any manifestation of extra dimensions would lead to a much improved 
understanding of the structure of space-time.
\item Information on GUT breaking, extra dimensions or $CP$ violation will help 
understand why there are three generations (if that's the case).
\end{itemize}

\section{Conclusions}

There is as yet no convincing evidence of physics beyond the standard model, and care 
always needs to be exercised when anomalies are seen at the edge of an experiment's 
sensitivity.  The good news, however, is that things would need to conspire for new
physics to escape detection at the LHC.  While the measurement of new physics parameters
will start at the LHC, a linear collider with sufficient center-of-mass energy will 
be critical to the precise understanding of the underlying physics.

Among models of new physics, only supersymmetry deals with the hierarchy problem, 
gauge coupling unification and EWSB, but it comes at a significant price.  It seems
likely that most of the really fundamental questions, concerning for example 
the nature of electric and color charges, will remain unanswered for 
quite a while longer.

\section*{Acknowledgments}

The author would like to thank the organizers for their invitation and 
a very stimulating and enjoyable conference.

\end{document}